\documentclass[twocolumn,preprintnumbers,msmath,amssymb,prl,floatfix,final]{revtex4}

\usepackage{graphicx}
\usepackage{dcolumn}
\usepackage{bm}
\usepackage[latin1]{inputenc}
\usepackage[latin1]{inputenc}
\usepackage{verbatim}
\usepackage{float}
\usepackage{color}

\usepackage[]{placeins}
\usepackage{slashbox}

\graphicspath{{./figures/}{./eps/}{./pdf/}}

\begin{document}

\title{Solution-based alkyne-azide coupling on functionalized Si(001) prepared under UHV conditions }

\author{T. Glaser$^{1}$, J.~Meinecke$^{2}$, C.~L\"anger$^{1}$, J.~Heep$^{1}$, U.~Koert$^{2}$,
and M.~D\"urr$^{1,*}$}
\address{
$^1$Institut f{\"u}r Angewandte Physik and Zentrum f\"ur Materialforschung, Justus-Liebig-Universit\"at Giessen, Heinrich-Buff-Ring 16, D-35392 Giessen, Germany\\
$^{2}$Fachbereich Chemie, Philipps-Universit\"at Marburg, Hans-Meerwein-Stra{\ss}e 4, D-35032 Marburg, Germany\\
$^{*}$E-mail-address: michael.duerr@ap.physik.uni-giessen.de\\}
\date{\today}

\bibliographystyle{prsty}
\hyphenation{tempera-ture}

\begin{abstract}
Synthesis of organic bi-layers on silicon was realized by a combination of surface functionalization under ultra-high vacuum (UHV) conditions and solution-based click chemistry. The silicon (001) surface was prepared with a high degree of perfection in UHV and functionalized via chemoselective adsorption of ethinyl cyclopropyl cyclooctyne from the gas phase. A second organic layer was then coupled in acetonitrile via the copper-catalyzed alkyne azide click reaction. The samples were directly transferred from UHV via the vapour phase of the solvent into the solution of reactants and back to UHV without contact to ambient conditions. Each reaction step was monitored by means of X-ray photoelectron spectroscopy in UHV; the N~1s spectra clearly indicated the click reaction of the azide group in the two test molecules employed, i.e., methyl-subsituted benzylazide and azide substituted pyrene. In both cases, up to 50 - 60~$\%$ of the ethinyl cyclopropyl cyclooctyne molecules on the surface were reacted.
\end{abstract}

\maketitle


\section*{INTRODUCTION}

Click reactions such as the most prominent alkyne azide coupling \cite{Kolb01AngChem,Sletten09AngChem} are generally performed in the presence of a catalyst dissolved in an adequate solvent. When solution-based click reactions are applied on surfaces \cite{Collman06langmuir,Devaraj07qcs,Haensch08Nanotechnology,Marrani08ElectroActa,Ciampi11Langmuir,Li11ChemAsian,Lo12Langmuir,Gouget-Laemmel12jpcc}, the initial functionalization of the surface prior to ''clicking`` the second layer is typically performed in solution as well. Whereas this is possible for passivated surfaces \cite{Haensch08Nanotechnology,Marrani08ElectroActa,Ciampi11Langmuir,Li11ChemAsian,Lo12Langmuir,Gouget-Laemmel12jpcc} and surfaces with an inherently low reactivity \cite{Collman06langmuir,Devaraj07qcs}, highly reactive surfaces such as pristine Si(001) are typically prepared and stored under UHV conditions in order to guarantee a high level of cleanliness and structural perfection. The direct application of solution-based click chemistry schemes to such vacuum-processed surfaces thus seems as an experimental contradiction. However, this combination could open the route to synthesizing new organic molecular architectures, e.g., with tailored optical or physico-chemical properties, within the framework of vacuum-based technologies.

%
\begin{figure}[b!]
	\begin{center}
		\includegraphics[width = 7.5cm]{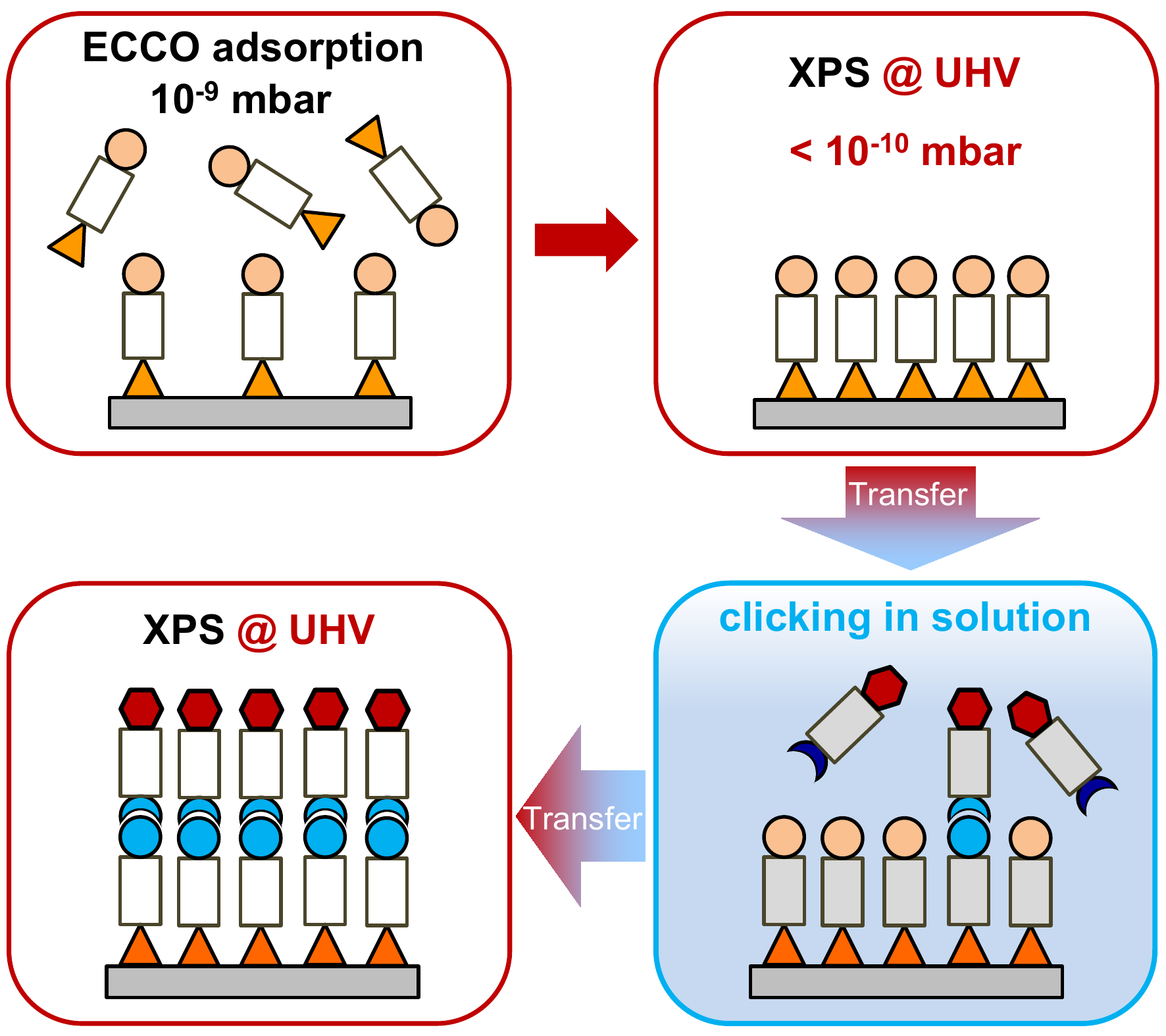}
		\caption[]{Schematic representation of the experiment. Top left: The bifunctional organic molecule ECCO is chemoselectively adsorbed on the bare Si(001) substrate via its strained triple bond (compare \textbf{1} and \textbf{2} in Fig.~\ref{fig:Scheme}). Top right: An XPS measurement is carried out under UHV conditions. Bottom right: After transferring the sample into the reaction chamber (acetonitrile vapour phase), the second organic layer is build on the surface using azide/alkyne click chemistry in solution. The latter is kept in a trough on the bottom of the reaction chamber (compare Fig.~\ref{fig:Kammer}(b))). Bottom left: After transfer back into the main chamber, the reaction product is analyzed by means of XPS.
			 \label{fig:Intro}}
	\end{center}
	\vspace{-5mm}
\end{figure}

Here we use surface functionalization based on substituted cyclooctynes and their chemoselective reactivity on Si(001) \cite{Reutzel16jpcc,Laenger19jpcm,Pecher18tca,Glaser20meeco} in combination with a direct transfer from UHV into solution in order to apply solution-based click chemistry on a silicon surface prepared and functionalized under UHV conditions. The experimental procedure is sketched in Fig.~\ref{fig:Intro}: Chemoselective adsorption of a monolayer of ethinyl cyclopropyl cyclooctyne (ECCO, \textbf{1} in Fig.~\ref{fig:Scheme}) from gas phase is followed by X-ray photoelectron spectroscopy (XPS) analysis and subsequent transfer of the functionalized surface into the reaction chamber. In the latter one, alkyne azide coupling in solution leading to triazole \textbf{4} in Fig.~\ref{fig:Scheme} is performed. After completion of the second layer, the sample is transferred back into the main chamber and the reaction products are again analyzed by means of XPS. Reaction yields in terms of the relative number of reacted ECCO molecules on the surface were measured for methyl-subsituted benzylazide (\textbf{3} in Figs.~\ref{fig:Scheme} and \ref{fig:Kammer}(a)) and azide functionalized pyrene (\textbf{5} in \ref{fig:Kammer}(a)) as a function of dose with and without additives to the copper catalyst.

\section*{EXPERIMENTAL METHODS }

%
\begin{figure}[t!]
	\begin{center}
		\includegraphics[width = 7.0cm]{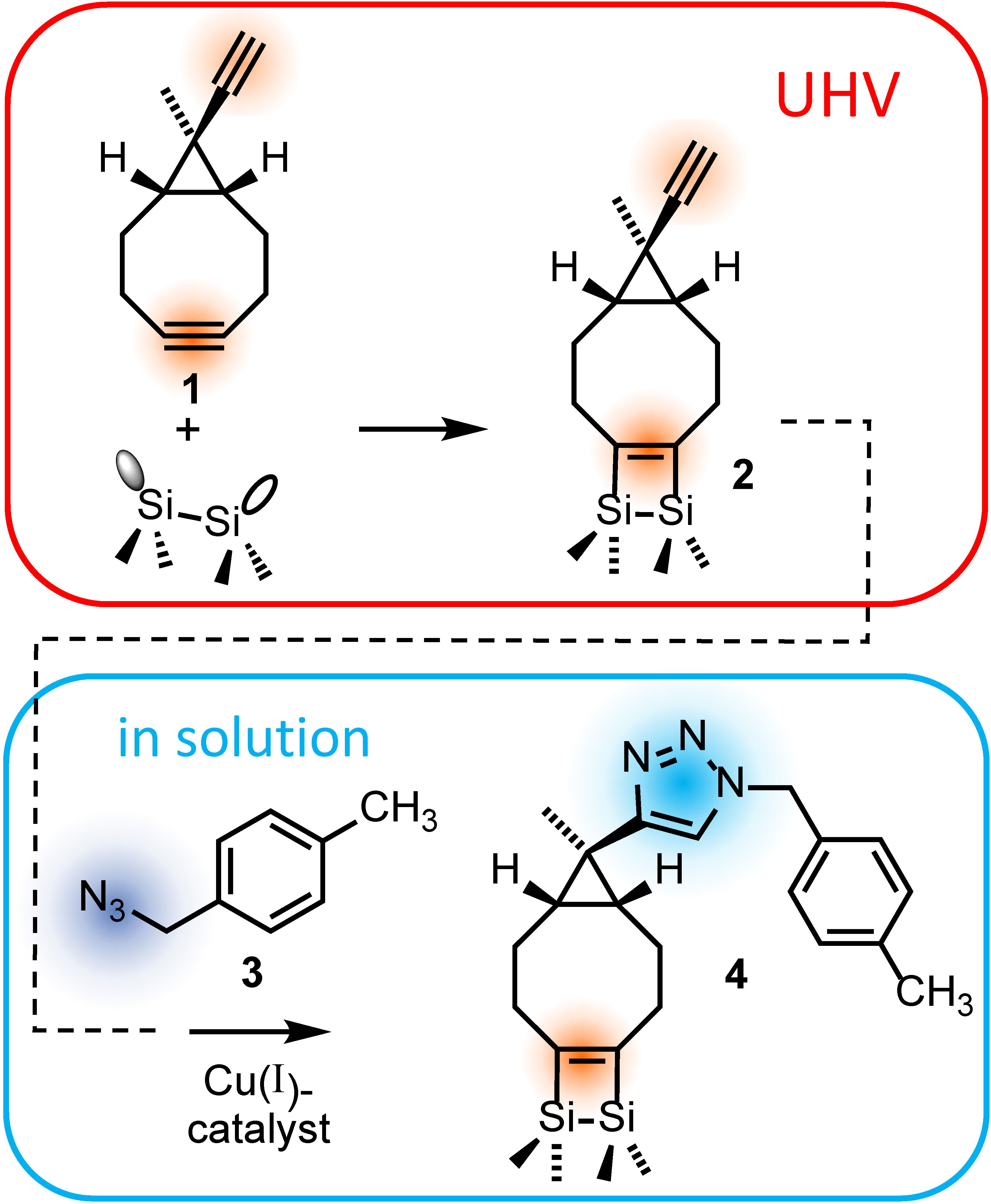}
		\caption[]{ ECCO (\textbf{1}) adsorbs chemoselectively via the strained triple bond on Si(001)(\textbf{2}) under UHV conditions \cite{Laenger19jpcm}. ''Clicking`` of methyl-substituted benzylazide (\textbf{3}) on ECCO-covered Si(001) using a catalyst such as CuI leads to the product \textbf{4}.
			 \label{fig:Scheme}}
	\end{center}
	\vspace{-5mm}
\end{figure}

Sample preparation was performed in a UHV chamber with a base pressure $<~1~\times~10^{-10}~$mbar. Si(001) samples were prepared by degassing at 700~K and repeated direct current heating cycles to 1450~K.
A well ordered $2~\times~1$ reconstruction was obtained by cooling rates of about 1~K$/$s \cite{Schwalb07prb,Mette09cpl}.

\begin{figure}[]
	\begin{center}
		\includegraphics[width = 8.5cm]{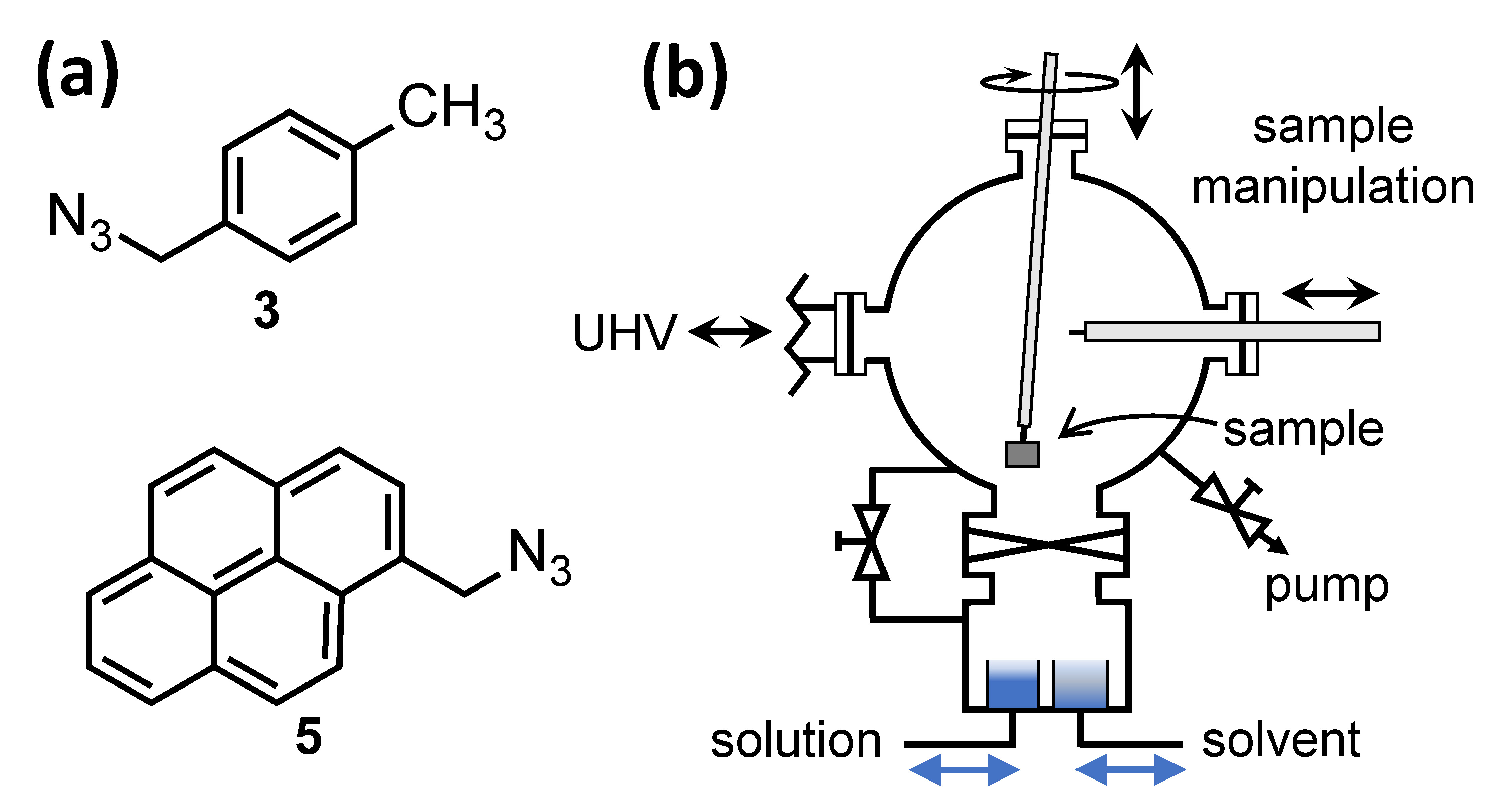}
		\caption[]{(a) Structural formulas of the of the reactants used in the click experiments, methyl-substituted benzylazide (\textbf{3}) and pyrene functionalized azide (\textbf{5}). (b) Schematic drawing of the separate reaction chamber. The samples are transferred via a sample transfer rod and an wobble stick into the stainless steel beakers fixed on the bottom of the chamber. They are filled with the azide solution and the solvent for rinsing. The main part of the reaction chamber can be separated from the two beakers by a UHV gate valve. The pressure in the two parts of the reaction chamber can be equalized when opening the valve in the bypass.
			 \label{fig:Kammer}}
	\end{center}
	\vspace{-5mm}
\end{figure}

Ethinyl cyclopropyl cyclooctyne (ECCO, \textbf{1} in Fig.~\ref{fig:Scheme})) was synthesized according to Ref.~\cite{Muenster16orglett}. Synthesis of methyl-subsituted benzyl azide (short: benzyl azide, \textbf{3} in Fig.~\ref{fig:Scheme}) was carried out via the corresponding benzyl bromide by nucleophilic substitution with sodium azide \cite{Song17OrgLett,Cilliers19cbdd}. Synthesis of 1-(azidomethyl)pyrene (short: pyrene azide, \textbf{5} in Fig.~\ref{fig:Kammer}) was carried out according to Ref.~\cite{vanderKnaap11ChemMedChem}. CuI and CuBr(PPh$_3$)$_3$ are commercially available. CuI(PPh$_3$)$_3$ was synthesized in one step from CuI and triphenylphosphine according to Ref.~\cite{Khan19NewJChem}. Both, CuBr(PPh$_3$)$_3$ and CuI(PPh$_3$)$_3$, were stored under inert conditions at room temperature until they were solved in dried acetonitrile. Typical azide concentrations in solution were 0.5 mol/L. The catalysts were added in a concentration of $5\cdot10^{-3}$~mol/L.

%

Following the procedure outlined in Fig.~\ref{fig:Intro}, the Si(001) surface was first functionalized with ECCO leading to \textbf{2} in Fig.~\ref{fig:Scheme} \cite{Laenger19jpcm}. The ECCO-covered sample was then transferred within 1 minute from the main chamber into the reaction chamber (Fig.~\ref{fig:Kammer}(b)) via an additional, separately pumped vacuum stage. The reaction chamber was fully baked out prior to the experiments ($p_{\rm RC,base} \le 10^{-8}$~mbar); during the experiments, the pressure is determined by the vapour pressure of the solvent used for solving the reactants and for rinsing the samples, in our case acetonitrile (MeCN), and the pumping speed of the system. The samples were transferred into the stainless steel beaker filled with the azide solution with the help of a wobble stick: the reaction time of the sample in solution was chosen between 4 to 60 minutes. After the reaction, the sample was rinsed in pure, dried acetonitrile which was kept in the second stainless steel beaker in the reaction chamber. Typical rinsing time was 60 minutes. After rinsing, the samples were transferred back into the main chamber, typical increase of pressure in the main chamber after transfer was below $\Delta p = 2 \times 10^{-10}$~mbar.

XPS measurements were performed in the main chamber using an Al~K$_{\alpha}$ X-ray source with a monochromator (Omicron XM1000) and a hemispherical energy analyzer (Omicron EA125). All XPS spectra were referenced to the Si~2p$_{3/2}$ peak at 99.4~eV \cite{Reutzel15jpcc}. Voigt-profiles were used for fitting the data; they were composed of Gaussian and Lorentzian contributions (90~\% and 10~\%, respectively). If not otherwise stated, full width at half maximum (FWHM) of the single components was approximately 0.9~eV in case of the C~1s and O~1s signals, and approximately 1.5~eV for the N~1s signals as applied previously for comparable systems measured in this setup \cite{Laenger18jpcc,Heep20jpcc}.

\section*{RESULTS and DISCUSSION}

%
\begin{figure}[b!]
	\begin{center}
		\includegraphics[width = 8.5cm]{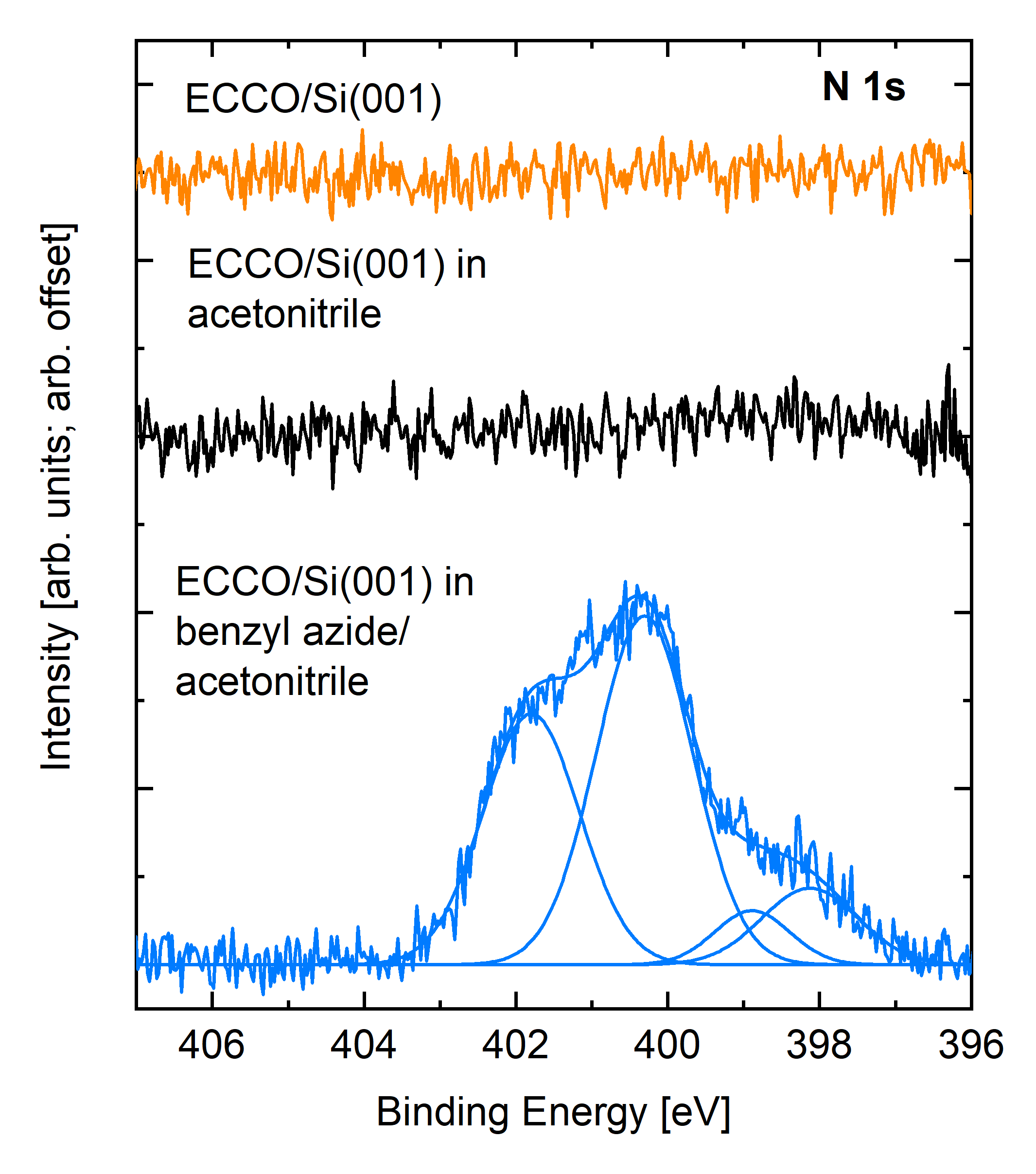}
		\caption[]{N~1s spectra taken after different steps of the experiment. Top: spectrum measured on the Si(001) sample after ECCO adsorption (orange). No peak is observed.
Center: The spectrum in black was measured after the ECCO covered Si(001) sample was kept in liquid acetonitrile for 15 minutes. No peak is observed, either. Bottom: The spectrum shown in blue was measured after reaction of the ECCO-covered Si(001) sample in azide solution. Four different peaks can be deduced between 397 and 403~eV, indicative of alkyne azide coupling. No signal is observed at 405~eV. 			 \label{fig:N1s-spectra}}
	\end{center}
	\vspace{-5mm}
\end{figure}

In Fig.~\ref{fig:N1s-spectra}, core level spectra in the N~1s energy region taken at different stages of the experiment, using benzyl azide as reactant, are compared. The data shown in orange in Fig.~\ref{fig:N1s-spectra} were measured on an ECCO-covered Si(001) surface. No signal is observed indicating that no nitrogen containing species were adsorbed neither on the bare nor on the ECCO-covered surface.
The spectrum taken after rinsing the ECCO-covered sample in pure acetonitrile is shown in black. No peaks are observed either, thus no nitrogen containing species were attached to the surface during transfer and rinsing in acetonitrile. In other words, the ECCO-covered Si(001) surface is not reacting with the chosen solvent, i.e., acetonitrile.
The spectrum shown in blue in Fig.~\ref{fig:N1s-spectra} was measured after the ECCO-covered Si(001) sample was transferred into the benzylazide/acetonitrile solution (60~min, CuBr(PPh$_3$)$_3$ catalyst). The spectrum can be decomposed into four different peaks: The peaks at the highest binding energy (401.9~eV and 400.5~eV) are assigned to the nitrogen atoms in the product of alkyne azide coupling (\textbf{4} in Fig.~\ref{fig:Scheme}), in good agreement with literature \cite{Collman06langmuir,Gouget-Laemmel12jpcc}. The peak at 400.5~eV is associated with the two N atoms bound to one N and one C atom (N-\textbf{N}-C), the peak at 401.9~eV is associated with the nitrogen atom in the N-\textbf{N}-N configuration. The intensity ratio differs from the stoichiometric ratio of 2:1, however, a small amount of copper catalyst remaining on the surface shifts a part of the N~1s intensity of the click product to lower binding energy, which influences the intensity ratio between the two components. In particular, the third peak at a binding energy of 399~eV can be attributed to the influence of the remainders of the catalyst. Upon heating the surface, the intensity of this peak is reduced and the intensity of the peak at 400.5~eV is increased along with a decrease of the catalysts' signal (see below). The peak at 398.1~eV can be assigned to N atoms directly bound to silicon \cite{Heep20jpcc}, indicating that a small amount of the azide molecules binds directly to the Si(001) surface. The intensity of this peak was comparable in all experiments, independent of the reaction time of the sample in solution. We thus conclude that a small amount of unreacted dangling bonds is still available on the ECCO-covered Si(001) surface which is readily reacted by the benzyl azide molecules. Most important, no peak at a binding energy of 405~eV is observed. The latter is assigned to the intact azide group \cite{Collman06langmuir,Gouget-Laemmel12jpcc} (compare Fig.~S1 in the Supporting Information), thus no intact benzyl azide molecules contribute to the N~1s signal intensity.

\begin{figure}[]
	\begin{center}
		\includegraphics[width = 8.5cm]{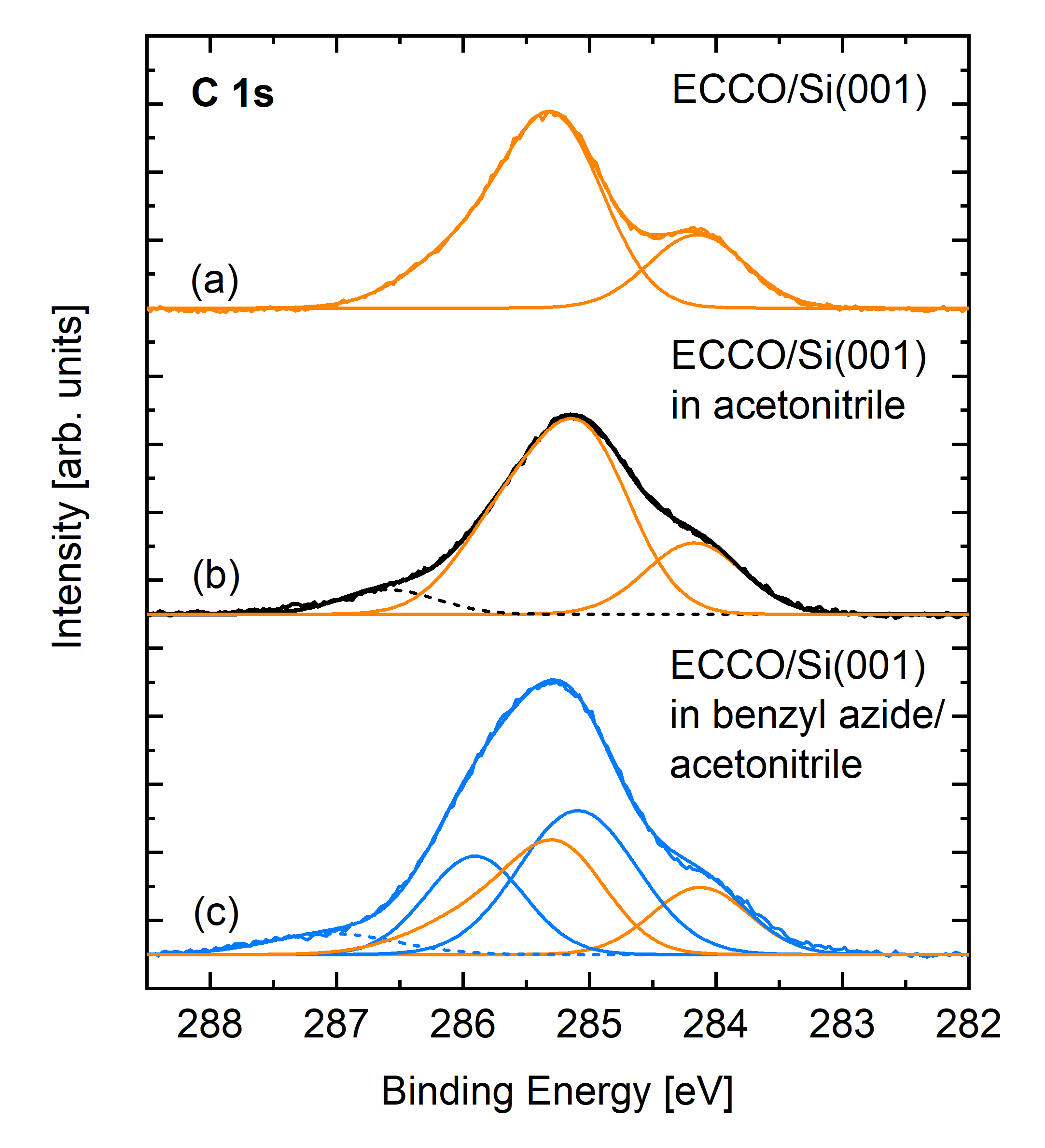}
		\caption[]{C~1s spectra after different reaction steps of the sample. (a) Spectrum taken on the Si(001) surface covered with ECCO molecules. Two major peaks can be observed, the peak around 285.3~eV represents the main body of the molecule and shows in total a greater width than the single components applied. (b) The spectrum in black represents the ECCO-covered Si(001) sample after 15~min in liquid acetonitrile. The spectrum is almost identical to the spectrum in (a), only a small additional peak towards higher binding energies is observed which is attributed to the influence the adsorption of oxygen-containing species. (c) Spectrum measured after the reaction of the ECCO-covered Si(001) sample in azide solution. In addition to the peaks attributed to the adsorbed ECCO molecules (orange, reduced intensity of the component at 285.2~eV when compared to (b) due to reaction of the terminal alkyne group), additional intensity is clearly observed. Two major additional components (blue solid lines) are fitted at 285.1 and 285.9~eV with a width of 1.1 and 0.9~eV, respectively.
			 \label{fig:C1s}}
	\end{center}
	\vspace{-5mm}
\end{figure}

The C~1s spectra shown in Fig.~\ref{fig:C1s} were obtained from the same experiments the O~1s spectra shown in Fig.~\ref{fig:N1s-spectra} were obtained from. The data shown in Fig.~\ref{fig:C1s}(a) were measured on the Si(001) surface directly after adsorption of ECCO. Two main components are observed, a stronger one around 285.3~eV and a smaller one at 284.2~eV. The former one, including the small tail to higher binding energies, indicates the body of C atoms bound to further carbon or hydrogen atoms, only; the latter one indicates carbon atoms covalently bound to the silicon surface \cite{Laenger19jpcm}. For ECCO, an intensity ratio of 5:1 between the two components indicates chemoselective adsorption via the strained triple bond (\textbf{2} in Fig.~\ref{fig:Scheme}), as discussed in detail in Ref.~\cite{Laenger19jpcm}. In the spectrum shown in Fig.~\ref{fig:C1s}(a), the ratio between the two components is rather 4:1 than 5:1, indicating that some of the ECCO molecules ($< 20$~\%) are attached both via the strained triple bond and the terminal triple bond. In Fig.~\ref{fig:C1s}(b), the spectrum measured after transferring the ECCO-covered sample into liquid acetonitrile is shown in black. The spectrum closely resembles the spectrum in Fig.~\ref{fig:C1s}(a), only an additional, small shoulder to higher binding energies is observed, which is attributed to the adsorption of some oxygen-containing species on the surface.
The spectrum shown in blue in Fig.~\ref{fig:C1s}(c) was measured after the ECCO-covered Si(001) sample was kept for 60 minutes in the benzylazide/acetonitrile solution (CuBr(PPh$_3$)$_3$ catalyst) and subsequent rinsing in acetonitrile. Compared to the ECCO spectrum (Figs.~\ref{fig:C1s}(a) and (b)), the total intensity of the signal is increased by about 60~\%. The additional intensity can be described with two further major components. The component at lower binding energy (285.1~eV) then represents the sum of carbon atoms in the main body of the methyl-substituted benzyl entity (C-\textbf{C}-C and C-\textbf{C}-H configurations), the component at higher binding energy (285.9~eV) represents carbon atoms in direct neighborhood of nitrogen atoms \cite{Heep20jpcc}, in agreement with alkyne azide coupling of the benzyl azide with the ECCO molecules adsorbed on Si(001). The intensity ratio between these two peaks is approx.~0.6, slightly higher than expected from the ratio of 3:7 for the two groups of C atoms in the triazole ring and in the methyl-substituted benzyl entity. Furthermore, the reduction of the intensity of the peak at 285.2~eV associated with the carbon atoms (C-\textbf{C}-C and C-\textbf{C}-H configurations) in the cyclooctyne unit is somewhat higher than expected from the reaction of the two C atoms of the terminal alkyne group. These differences might be attributed to (i) a more complex distribution of the peak intensity on the single components and (ii) some contribution from contaminations containing additional carbon atoms. In total, however, the C~1s spectrum in Fig.~\ref{fig:C1s}(c) fits well to the results obtained from the N~1s spectra in Fig.~\ref{fig:N1s-spectra}.

The spectra obtained with pyrene azide are qualitatively very similar to the spectra shown in Figs.~\ref{fig:N1s-spectra} and \ref{fig:C1s}, in particular, the N~1s spectra are almost identical, indicating the same coupling product. Only the increase in intensity in the C~1s intensity is in general higher for pyrene azide when compared to benzyl azide due to the higher number of carbon atoms in pyrene.

\begin{figure}[t!]
	\begin{center}
		\includegraphics[width = 8.5cm]{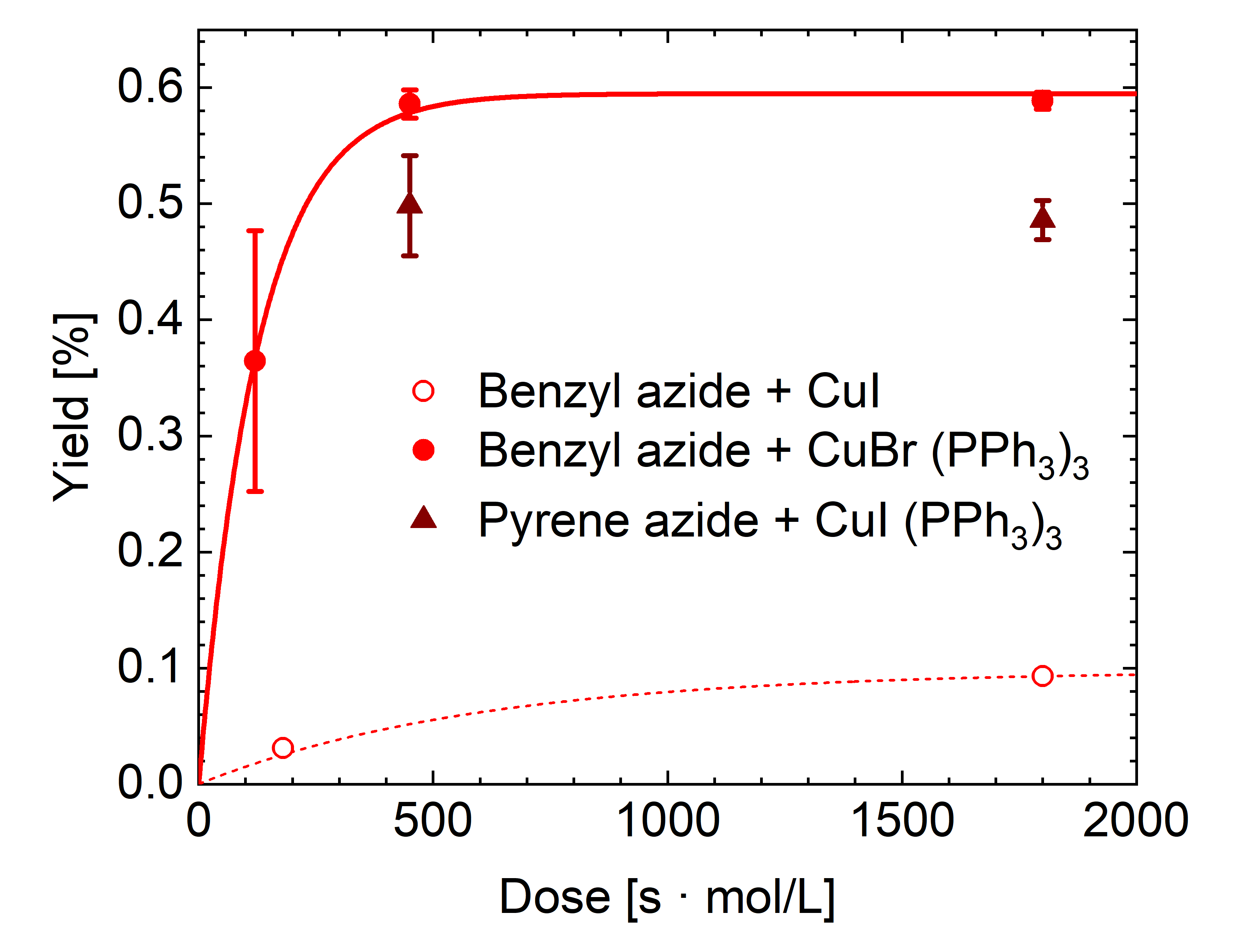}
		\caption[]{Yield of alkyne azide coupling as a function of dose for the molecules and Cu(I) catalysts investigated. For the pure CuI catalyst, no higher yield than approx. 10~$\%$ was observed (open red circles). With triphenyl phosphine added to the Cu(I)-catalyst, a yield up to 60~$\%$ was observed both for benzyl azide with CuBr (red dots) and pyrene azide with CuI. Lines are guides to the eye. All experiments were performed with a concentrations of the azide solution of 0.5~mol/L, except for the low dose experiment with CuI (0.05~mol/L, left open red circle).
			 \label{fig:yield}}
	\end{center}
	\vspace{-5mm}
\end{figure}

From the sensitivity-corrected intensities of the N~1s signal after click reaction and the C~1s signal of the ECCO-covered surface, the reaction yield in terms of reacted azide molecules per ECCO molecules on the surface was deduced. The results are summarized in Fig.~\ref{fig:yield} for three different systems; for the N~1s intensity, the component at 398~eV, which is attributed to nitrogen directly adsorbed on silicon, was not taken into account. Whereas a moderate yield of 10~\% was obtained with CuI as catalyst only, both CuI and CuBr lead to up to 60~\% yield in the presence of (PPh$_3$)$_3$. Pyrene azide shows a somewhat lower yield which is attributed to the larger molecules and steric constraints on the surface. Steric constraints might be also the reason why the yield is significantly below one in the case of benzyl azide, as well. However, one has to keep in mind that some of the adsorbed ECCO molecules (approx.\ 20~\%) are tethered with both functionalities and are thus not available for alkyne azide coupling. As a consequence, 75~\% of the {\sl available} ECCO molecules were reacted. The high reactivity is further indicated by the low dose needed for saturation of the click products, which translates into a reaction time of 15 minutes at an azide concentration of 0.5~mol/L.

In addition to the main constituents of the reactants, i.e., carbon and nitrogen, we observe further elements on the surface after the solution-based click reaction. First, small traces of I and Br are observed (3~\% and 2~\% with respect to adsorbed ECCO molecules, respectively). The Cu concentration amounts to  25~\%. These contaminants apparently originate from the catalysts used and their concentration on the surface might be further reduced when using a more elaborated rinsing scheme. Second, we detect a higher concentration of oxygen on the surface (approx.\ one oxygen atom on two ECCO molecules), both, after the full reaction cycle as well as when the sample is only rinsed in the pure solvent. On the other hand, the transfer process as such does not lead to a substantial oxygen signal. This points towards a contamination of acetonitrile with O$_2$/H$_2$O (Fig.~S2, Supporting Information) which might be further reduced by bubbling the solvent with inert gas and/or additional drying steps integrated in the manifold of our reaction chamber.

\section*{CONCLUSION}

In conclusion, we could demonstrate that solution-based click chemistry can be applied to highly reactive surfaces which have been prepared and functionalized under UHV conditions. Using Cu(I) catalysts in the presence of (PPh$_3$)$_3$, alkyne azide coupling was observed with high yield on the ECCO-functionalized Si(001) surface. The strategy is based on the chemoselective adsorption of bifunctional cyclooctynes on Si(001), which serves as a versatile interface between the inorganic semiconducor and the world of organic chemistry. Using a pyrene substituted azide as reactant for the alkyne azide coupling of the second layer, an optically active molecular structure was covalently attached to silicon in a controlled manner. This demonstrates the potential of our strategy for a wide range of applications.

\section*{Associated Content}

In the Supporting Information, N~1s and C~1s spectra of the benzyl azide adsorbed at 150~K on the ECCO-covered Si(001) surface from the gas phase and O~1s spectra taken after click reaction in solution are shown.

\section*{Acknowledgement}

We acknowledge financial support by the Deutsche Forschungsgemeinschaft through SFB~1083 (project-ID 223848855).

\bibliographystyle{prsty}

\bibliography{OrgMolSi,misc2,Lit_V2}

\end{document}


\renewcommand{\thepage}{S\arabic{page}}
	\renewcommand{\thefigure}{S\arabic{figure}}
	\renewcommand{\thetable}{S\arabic{table}}

	\setstretch{1.6}
	\noindent Supporting Information for:
	\begin{center}
		\begin{large}
			\textbf{Solution-based alkyne-azide coupling on functionalized Si(001) prepared under UHV conditions}\\
		\end{large}
		\vspace{3mm}
		\begin{small}
		T. Glaser$^{1}$, J.~Meinecke$^{2}$, C.~L\"anger$^{1}$, J.~Heep$^{1}$, U.~Koert$^{2}$,
and M.~D\"urr$^{1,*}$\\
		
$^1$\emph{Institut f\"ur Angewandte Physik and Zentrum f\"ur Materialforschung,
Justus-Liebig-Universit\"at Giessen, Heinrich-Buff-Ring 16, D-35392 Giessen, Germany}\\
$^2$\emph{Fachbereich Chemie, Philipps-Universit\"at Marburg, Hans-Meerwein-Stra{\ss}e 4, 35032 Marburg, Germany}\\
$^*$\emph{Corresponding author: michael.duerr@ap.physik.uni-giessen.de}
		\end{small}

		\date{\today}

	\end{center}
	\setstretch{1.3}
	\vspace{30mm}
	This Supporting Information includes
	\begin{itemize}
		\item[\textbf{(I)}] N~1s and C~1s spectra of a multilayer of benzyl azide on ECCO-covered Si(001)
        \item[\textbf{(II)}] O~1s spectra after reaction of pyrene azide on ECCO-covered Si(001)

		\end{itemize}
	
	\pagebreak

\noindent \textbf{I. N~1s and C~1s spectra of a multilayer of benzyl azide on ECCO-covered Si(001)}\\

\begin{figure}[h!]
	\begin{center}
		\includegraphics[width = 12cm]{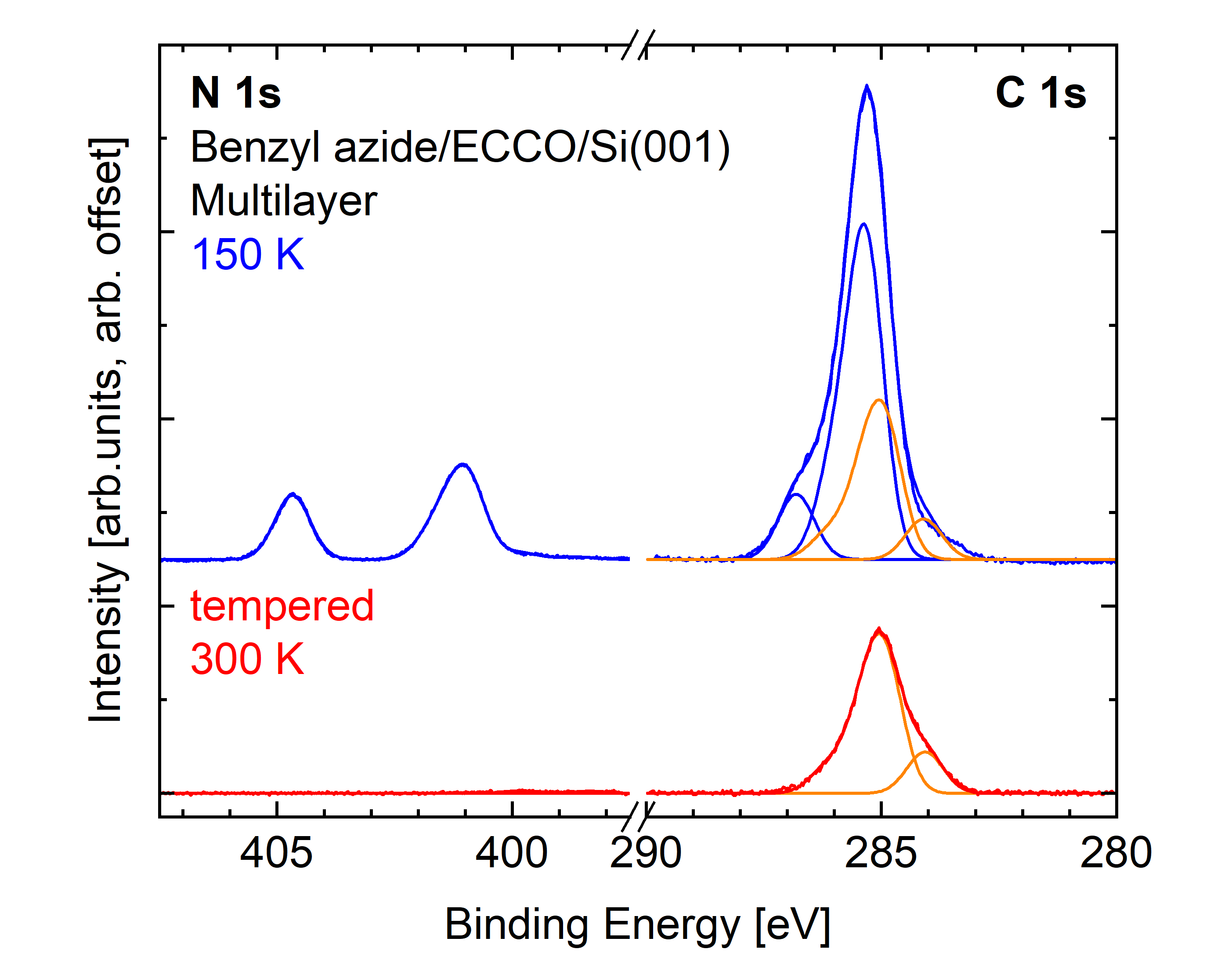}
		\caption[]{N~1s and C~1s spectra taken after gas phase adsorption of a multilayer of benzyl azide on ECCO-covered Si(001) at 150~K (top, blue). 2 peaks are observed in the N~1s spectrum at 404.7~eV and 401.0~eV (intensity ratio approx.~1:2), indicating the intact azide group of the benzyl azide physisorbed on the ECCO-covered Si(001) surface. In the C~1s spectrum, the orange components  are assigned to ECCO adsorbed on Si(001) underneath the multilayer (compare red spectrum taken after heating the sample to 300~K). The blue components at 285.3~eV and 286.8~eV are attributed to the C atoms in the benzyl azide molecules \cite{Heep20jpcc}. After heating the sample to 300~K, no peaks are observed in the N~1s spectrum, indicating that no azide is bound covalently to the surface.
			 \label{N1s_C1s}}
	\end{center}
	\vspace{-5mm}
\end{figure}

\newpage

\noindent \textbf{II. O~1s spectra after reaction of pyrene azide on ECCO-covered Si(001) }\\
	\\
	
\begin{figure}[h!]
		\begin{center}
		\includegraphics[width = 12cm]{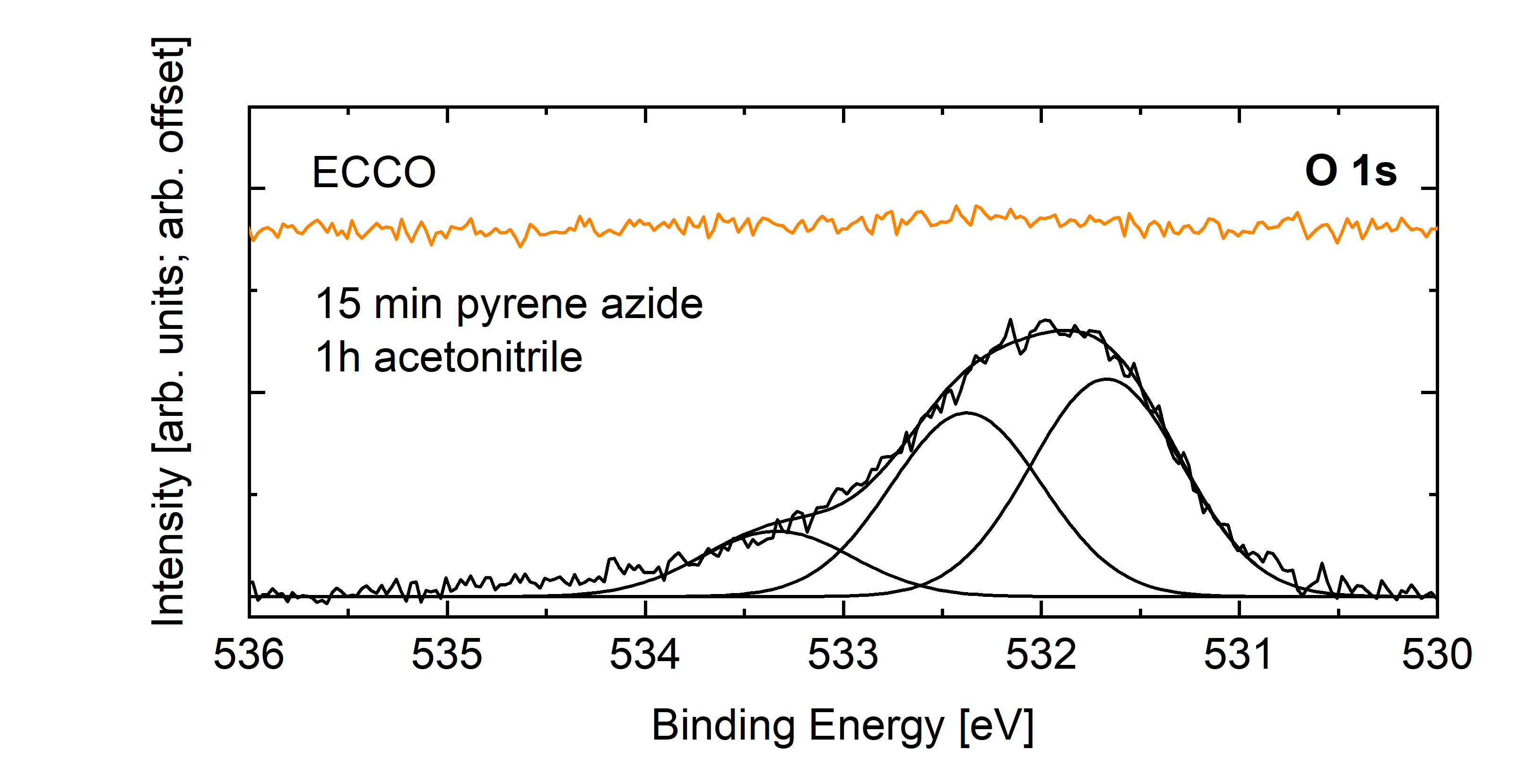}
		\caption[]{O~1s spectra of ECCO on Si(001) (top, orange) and of the ECCO-covered Si(001) surface after click reaction with pyrene azide in acetonitrile (bottom, black). 3 different peaks are observed at 531.7~eV, 532.4~eV, and 533.3~eV.}
		\label{O1s}
		\end{center}
	\end{figure}

In Fig.~\ref{O1s}, two O~1s spectra are shown. The spectrum on top (orange) represents the measurement after ECCO adsorption on Si(001). No peak is observed, thus no oxygen is bound to the Si(001) surface, neither directly nor via a molecular linker.
The second spectrum, shown on the bottom in black, was measured after the click reaction in solution. Three peaks can be deduced.
The peak at the lowest binding energy of 531.7~eV is assigned to oxygen binding with two Si-surface atoms, forming Si-\textbf{O}-Si \cite{Hofer85prl, Hofer89prb, Hofer89ss}. It might origin from the reaction of molecular oxygen \cite{Hofer85prl, Hofer89prb, Hofer89ss}.
The peak at a binding energy of 532.4~eV is assigned to oxygen atoms bound to a single silicon atom forming Si-\textbf{O}-R, where R can be either a hydrogen atom or a hydro-carbon entity \cite{Mette14, Reutzel15jpcc, Laenger18jpcc}. In the former case, the signal might be attributed to small amounts of water in the solvent. The peak at the highest binding energy (533.3~eV) is typically assigned to oxygen atoms in a chemical environment containing carbon and hydrogen atoms only, i.e., R-\textbf{O}-R, as it is found, e.g., in ethers \cite{Reutzel16jpcc} or alcohols.

\bibliographystyle{prsty}

\bibliography{OrgMolSi,misc2,Lit_V2}